\documentclass{optica-article}
\journal{opticajournal} 
\articletype{Research Article}
\pdfoutput=1

\usepackage{lineno}

\usepackage{upgreek}
\usepackage[capitalise]{cleveref}
\usepackage{chemformula}

\usepackage[textsize=small, 
tickmarkheight=5pt, 
color=blue!30, 
bordercolor=blue!30]{todonotes}

\begin{document}

\title{Tunable Degenerate Optical Parametric Oscillation with Coupled Microresonators}

\author{Nathalia B. Tomazio\authormark{*,1,2,3}, Luca O. Trinchão\authormark{*,1,2}, Eduardo S. Gonçalves\authormark{*,1,2}, Laís Fujii dos Santos \authormark{*,1,4}, Paulo F. Jarschel\authormark{1,2}, Felipe G. S. Santos\authormark{1,2}, Thiago P. Mayer Alegre\authormark{1,2}, Gustavo S. Wiederhecker\authormark{1,2,$\dag$}}

\address{
\authormark{*} These authors contributed equally to this work.\\
\authormark{1} Photonics Research Center, Universidade Estadual de Campinas (UNICAMP), Campinas, SP, Brazil\\
\authormark{2} Gleb Wataghin Physics Institute, Universidade Estadual de Campinas (UNICAMP), Campinas, SP, Brazil
\authormark{3} Instituto de Física, Universidade de São Paulo, São Paulo, Brazil\\
\authormark{4} School of Electrical Engineering and Computer Science, University of Ottawa, K1N 6N5, ON, Canada\\
}

\email{\authormark{$\dag$}gsw@unicamp.br}

\date{\today}

\begin{abstract*}
Microresonator-based degenerate optical parametric oscillation (DOPO) has recently been explored as a compelling platform for all-optical computing and quantum information applications, such as truly random number generation and the production of squeezed states of light. Emerging research has highlighted the potential of coupled microresonators, or photonic molecules, as a novel avenue for spectral engineering, unlocking an extra degree of freedom for the optimization of four-wave mixing interactions. Here, we demonstrate DOPO within the coupled modes of a silicon nitride triple-state photonic molecule. 
Our design introduces a distinctive mechanism for spectral engineering, using microheaters to individually tune the resonance spectral positions, thus enabling dynamic local dispersion control within the coupled modes. We successfully generate a DOPO signal with active efficiency control and explore the optical mode spacing in the tens of gigahertz range to use native phase-locked optical pumps driven by a radio-frequency source. 

\end{abstract*}

\section{Introduction}

Degenerate optical parametric oscillation (DOPO) has been investigated as a novel strategy for various applications, including all-optical quantum random number generation (RNG)~\cite{marandi2012all,okawachi2016quantum,okawachi2021dynamic}, generation of squeezed states of light~\cite{zhao2020near, zhang2021squeezed} and coherent optical computing~\cite{marandi2014network, okawachi2020demonstration, inagaki2016large}. 
DOPO occurs when the signal and idler fields generated by parametric amplification in an optical cavity are degenerate in frequency. 
In $\chi^{(2)}$ and $\chi^{(3)}$-media, DOPO is generated, respectively, via parametric down-conversion and four-wave mixing (FWM) with a dual-pump scheme. In both cases, the DOPO phase properties arise from the underlying phase-sensitive parametric amplification~\cite{mckinstrie2004phasesensitive, marandi2012all}. 
Below the oscillation threshold, the phase state of the signal and idler pair is described by a squeezed vacuum state~\cite{zhao2020near, zhang2021squeezed}. 
Upon reaching the oscillation threshold -- where parametric gain exceeds losses -- DOPO undergoes a nonequilibrium phase transition to a coherent binary-phase state, which is either in-phase or $\pi$ out-of-phase relative to the pump fields~\cite{marandi2012all, nabors1990coherence}. 
In this case, both phase states are equally likely, since oscillation is initiated from vacuum fluctuations. 
This random binary phase state provides a means to truly RNG \cite{marandi2012all} and is used to emulate spin states on photonic Ising machines~\cite{inagaki2016large}.
Recently, CMOS-compatible integrated microresonators based on the $\chi^{(3)}$ nonlinearity have enabled the development of on-chip DOPO-based applications~\cite{okawachi2015dual, okawachi2016quantum, okawachi2021dynamic, okawachi2020demonstration, zhao2020near, zhang2021squeezed}, offering scalability, compactness, enhanced dispersion control, and high Q-factors, enabling efficient nonlinear effects even at low-power CW pumping.

Near-field coupled microresonators, also known as photonic molecules, offer extra degrees of freedom for spectral engineering\cite{espinel2017brillouin, little1997microring, popovic2006coupling, zengDesign2014, souza2015spectral, tusnin2023nonlinear, pal2024linear, singh2024efficient, wang2023integrated}.
These systems exploit evanescent coupling to induce hybridization of the bare resonator modes, resulting in frequency-split coupled modes (supermodes) that locally modify the dispersion.
By coupling two resonators, several authors have harnessed controllable avoided mode-crossings to engineer phase-matching conditions for FWM, allowing generation of optical frequency combs in the normal dispersion regime~\cite{xue_normal-dispersion_2015, fujii_analysis_2018, kimTurnkey2019}, purity enhancement and preservation of squeezed states~\cite{zhang2021squeezed}, and high-efficiency parametric conversion~\cite{gentry2014tunable}.
Despite significant progress in this direction, to the best of our knowledge, no nanophotonic device has yet demonstrated simultaneous tunability for dispersion control and isolation of the target FWM interaction from competing processes.
Additionally, for dual-pumped FWM interactions, such as DOPO, the participating modes of single or two coupled resonator platforms are spaced by free spectral range (FSR) units, ranging from tenths to units of THz. This wide frequency spacing inevitably demands two lasers for the dual-pump scheme, which can be experimentally challenging, especially for those applications requiring phase-locked pumps.

\begin{figure*}[t]
    \centering
    \includegraphics[width=\linewidth]{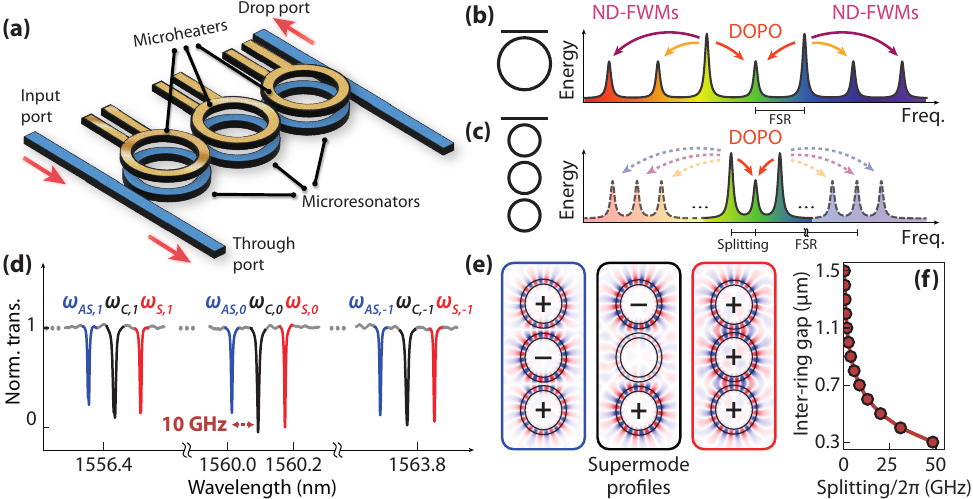}
    \caption{\textbf{Photonic molecule design:} 
    (a) Photonic molecule design comprising an array of three identical \ch{Si3N4} ring resonators with $50~\upmu$m of radii (FSR $\sim450$~GHz), a cross-section of $1~\upmu$m\,$\times\,0.8~\upmu$m (width $\times$ height) and an inter-ring gap of 550 nm, cladded in a SiO$_2$ substrate. Light is launched at the coupled microrings through the bus waveguide (waveguide-to-ring gap of 600~nm) and can be collected via either the through or the drop port. Inverted tapers at both the input and output ports optimize the TE coupling. Integrated microheaters atop each ring allow for spectral tunability.
    (b,c) Third-order parametric nonlinear processes that occur in a dual-pumped (a) single-ring microcavity and (b) a three-coupled ring photonic molecule. When phase-matching is achieved, DOPO is generated at the central resonance between the two pumps. In (b), the normal dispersion regime hinders the undesired ND-FWM interactions by phase-mismatching supermodes at different mode numbers.
    (d) Measured linear transmission at the through-port of the photonic molecule, exhibiting evenly spaced triplets occurring every free-spectral range of the bare rings. The colored Lorentzian dips indicate the Antisymmetric  (AS, blue), Central (C, black), and Symmetric (S, red) supermodes. The loaded quality factor of the C supermode was measured to be roughly $Q_L = 2.2 \times 10^5$.
    (e) 2D simulation of the electric field profile of each supermode using the finite-element method. This simulation uses a representative ring radius for better visualization.
    (f) Frequency splitting of supermodes as a function of the inter-ring gap for identical rings, calculated via CMT~\cite{haus1991coupled}. The line serves as a guide for the eye.
    }
    \label{fig:photonic-molecule}
\end{figure*}

In this work, we propose an array of three identical coupled SiN microresonators as a platform for DOPO, illustrated in \cref{fig:photonic-molecule}(a).
The mode hybridization induced by the coupling yields a triplet of coupled modes that repeats every FSR of the bare rings, so DOPO can be fully encompassed within one triplet of coupled modes without relying on modes of other azimuthal orders.
For comparison, in single-resonator DOPOs, achieving phase-matching for the three interacting modes often inadvertently phase-matches other azimuthal modes as well, leading to parasitic and cascading non-degenerate FWM (ND-FWM) processes, as illustrated in \cref{fig:photonic-molecule}(b). In contrast, our photonic molecule design can potentially isolate the DOPO from these competing effects. This is due to the wide spectral separation between triplets, allowing local phase-matching within a triplet to occur independently from phase-matching across other triplets (\cref{fig:photonic-molecule}(c)).
By tuning the individual phases of the coupled modes with integrated microheaters, we unlock locally tunable dispersion control, enabling us to dynamically correct for phase mismatch and optimize third-order nonlinear optical interactions within a triplet of coupled modes.
Moreover, by setting up inter-resonator gaps of a few hundred nanometers, frequency splittings of tens of GHz can be achieved among the coupled modes, without increasing cavity length~\cite{barea2013silicon, tikan2021emergent}.
Such radio-frequency (RF) range splittings allow for phase-coherent dual-tone pumping from a single laser source using electro-optic modulation (EOM).
This strategy confers native phase-locking between the pump fields, considerably reducing the complexity of the experimental setup for DOPO.

\section{Design considerations}

Our photonic molecule comprises an array of three identical (degenerate) SiN microring resonators, fabricated by Ligentec SA, illustrated in \cref{fig:photonic-molecule}(a). Near-field evanescent coupling between the rings induces hybridization of their individual modes, resulting in coupled modes that undergo frequency splitting from their original uncoupled frequencies. This results in evenly spaced triplets occurring in every FSR of the bare rings, shown in the through-port transmission in \cref{fig:photonic-molecule}(d). Moreover, integrated microheaters placed atop each microring enable dynamic tuning of the individual resonant frequencies via the thermo-optic effect. This capability allows for precise control over the coupling dynamics, enabling us to spectrally engineer the supermode splittings and locally tailor the group velocity dispersion (GVD) within a set of supermodes.

\begin{figure*}[t]
    \centering
    \includegraphics[width=\linewidth]{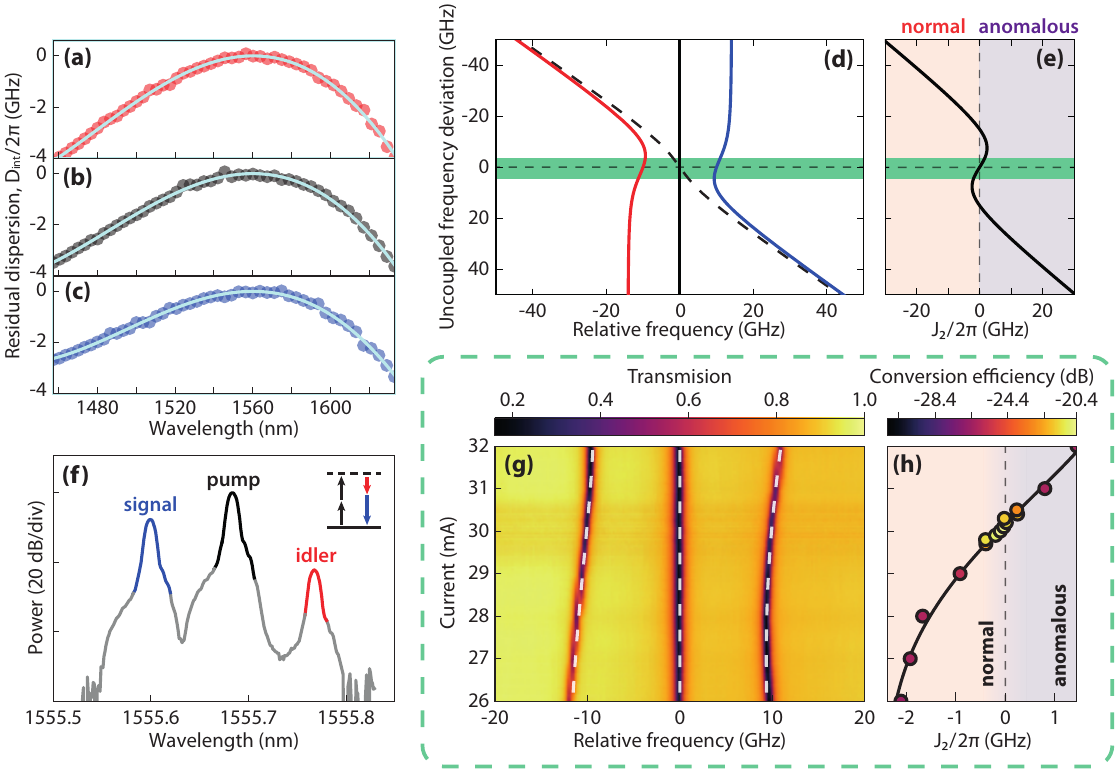}
    \caption{\textbf{GVD control and FWM:} 
    (a-c) Residual dispersion ($D_{int}=\omega_{j,\mu}-\omega_{j, 0} - D_{1,j}\mu$) for the S (a), C (b), and AS (c) supermodes, revealing their normal GVD regimes. Circles represent measured dispersion; line curves are cubic fittings.
    (d) Theoretical calculation of the S (red), C (black), and AS (blue) supermodes relative frequencies as a function of a frequency deviation of ring (1). The diagonal dashed curve represents the relative frequency deviation $(\delta\omega - \omega_\mathrm{C})/2\pi$. 
    (e) sGVD ($J_2$) calculated from (d), indicating the transition from the normal to the anomalous local supermode dispersion regime. The green hatched region in both (d, e) delimits the parameter region explored in our experiments depicted in (g, h).
    (f) Optical spectrum of FWM. The involved supermodes are colored following the color convention (S, red; C, black; AS, blue), while the insert indicates the energy conservation of the nonlinear process. A signal-to-idler conversion efficiency of -20.4 dB is displayed. 
    (g) Relative spectral position of supermodes collected at the through port as a function of the current applied to the microheater atop ring (1). The white dashed lines represent our theoretical model fitted to the experimental data. 
    (h) sGVD ($J_2$) calculated from (g), indicating the transition from the normal to the anomalous local supermode dispersion regime. The black solid curve indicates the fitted model of our calculation in (e) and the color-coded scatter points indicate the idler field power achieved in the stimulated FWM experiment.
    }
    \label{fig:dispersion-control}
\end{figure*} 

The supermode frequency splitting can be determined by solving the coupled amplitude basis $\{a_j\}$ and its associated eigenfrequencies $\omega_j$ via coupled mode theory in time (CMT)~\cite{haus1991coupled}: 
\begin{subequations}
	\begin{align}
	&a_\mathrm{AS} = \frac{1}{2}(1~-\sqrt{2}~1)^T, & &\omega_{\mathrm{AS}} = \omega_0 + \sqrt{2}J, &  \\
	&a_\mathrm{C} = \frac{1}{\sqrt{2}}(-1~0~1)^T, & &\omega_\mathrm{C} = \omega_0, &  \\
	&a_\mathrm{S} = \frac{1}{2}(1~\sqrt{2}~1)^T, & &\omega_\mathrm{S} = \omega_0 - \sqrt{2}J. & 
	\end{align}
    \label{eq:CMT}
\end{subequations}
The supermode amplitudes are normalized such that $|a_j|^2$ corresponds to the $j$-mode energy, for which $j = $ AS, C and S refer to the antisymmetric, central, and symmetric supermodes, respectively.
The electric field distribution of the supermodes, simulated by the finite element method (COMSOL Multiphysics), is depicted in \cref{fig:photonic-molecule}(e).
In the $a_\mathrm{AS}$ and $a_\mathrm{A}$ supermodes, the electric field displays opposite phases within the central resonator, which leads to their designation as antisymmetric  (AS) and symmetric (S) modes: in the AS supermode, the electric field in the central microring is out-of-phase with respect to the outer counterparts, while in the S supermode, the electric field in the three microrings is in phase. The central supermode $a_\mathrm{C}$ mostly occupies the outer rings and undergoes minimal frequency shifts due to coupling~\cite{popovic2006coupling}.
Such modal distributions have been shown to affect the FWM dynamics in coupled resonator systems~\cite {zeng2014design} (see Supplementary Material \cref{sec:SM_DOPOinmolecules} for details). 
Moreover, these supermodes deviate from the original uncoupled degenerate resonance frequency $\omega_0$ by the factor $\sqrt{2}J$, where $J/2\pi$ represents the ring-to-ring coupling rate, adjustable via the inter-ring gap. 
The frequency splitting between the supermodes is presented in \cref{fig:photonic-molecule}(f) as a function of the inter-ring gap distance. 
Our devices exhibit an inter-ring gap of 550 nm, which yields a radio-frequency (RF) range splitting ($\sqrt{2}J \sim 10~\text{GHz}$). The calculation details of the inter-ring coupling coefficient can be found in the Supplementary Material \cref{sec:SM_Inter-ring coupling}.

The supermodes' rich dispersive dynamics can be described by the resonator optical frequency expansion~\cite{liu2016Frequencycombassisted},
\begin{equation}
    \omega_{j, \mu} = \omega_{j, 0} + D_{1,j}\mu + \frac{1}{2}D_{2,j}\mu^2 + \dots
    \label{eq:freq-expansion}
\end{equation}
where $\mu = m - m_0$ is the azimuthal mode number relative to a reference mode ($m_0$) with frequency $\omega_{j, 0}$, and the coefficients $D_{1,j}$, $D_{2,j}$ represent the FSR and the change in the FSR with respect to $\mu$ (GVD parameter), respectively. The index $j$ is included to account for the supermode numbers ($j$ = AS, C, or S). Our device exhibits normal GVD regimes ($D_{2,j}<0$) as shown in \cref{fig:dispersion-control}(a-c), with $D_2/2\pi = -8.3$ (S), $-7.6$ (C) and $-6.7$ (AS) MHz, at $\lambda_0 \approx 1560$~nm. The GVD varies slightly for each supermode and higher-order dispersion terms have been neglected. 
For a given triplet (fixed mode number $\mu=\mu_0$), the supermode dispersion is described by defining the supermode free spectral range (sFSR, $J_{1}$) as the average frequency splitting and the supermode group velocity dispersion (sGVD, $J_{2}$) as its variation within the triplet, similar to standard FSR and GVD definitions across mode numbers~\cite{tusnin2023nonlinear} (for details on this treatment, see Supplementary Information \cref{sec:sm_supermode dispersion}).
In triple-state photonic molecules, up to second-order supermode dispersion is defined: the $J_{2}$ sign indicates whether the supermodes' frequency spacing increases ($J_2>0$) or decreases ($J_2<0$) with frequency, describing an anomalous and normal local dispersion regime, respectively. It can be calculated straightforwardly as $J_2 \equiv (\omega_\mathrm{AS} - \omega_\mathrm{C}) - (\omega_\mathrm{C} - \omega_\mathrm{S})$.

While \cref{eq:CMT} holds for the case of three identical bare rings ($\omega^{(1)} = \omega^{(2)} = \omega^{(3)} = \omega_0$), individual frequency deviations from the degenerate case can produce large spectral asymmetry among the triplet of supermodes, modifying the coupled basis solution. 
In \cref{fig:dispersion-control}(d), we show the supermode frequencies as a function of a deviation $\delta\omega$ in the eigenfrequency of the ring coupled to the bus waveguide, hereafter referred to as ring (1) ($\omega^{(1)} = \omega_0 + \delta\omega$, $\omega^{(2)} = \omega^{(3)} = \omega_0$). 
For $\delta\omega = 0$, the supermodes are evenly spaced, as shown in the representative spectrum depicted in \cref{fig:photonic-molecule}(d)). As $\delta\omega$ increases, the eigenfrequencies shift, resulting in the avoided mode crossing curves observed in \cref{fig:dispersion-control}(d). These shifts in the supermode frequencies can be exploited as a local dispersion control mechanism within a triplet of supermodes~\cite{herr2014mode}. We show in \cref{fig:dispersion-control}(e) that CMT predicts that the supermode frequency spacing $J_{2}$ is controllable and can reach a normal, zero, and anomalous local dispersion (sGVD). 
We experimentally demonstrate this local dispersion control by employing integrated microheaters atop each microresonator, allowing for controllable phase-shifting of the bare-ring resonances through the thermo-optic effect.
In \cref{fig:dispersion-control}(g), we illustrate the tuning of the relative frequency spacing of the supermodes achieved by varying the current supplied to the microheater on top of the ring (1). 
This enables active supermode dispersion engineering in our devices, allowing us to set up normal ($J_2<0$), zero ($J_2=0$), and anomalous ($J_2>0$) local dispersion regimes, as depicted in \cref{fig:dispersion-control}(h). 
Notably, this local dispersion control (sGVD) was found to have a negligible impact on the overall GVD of the supermodes.

To illustrate the control on sGVD achieved in our devices, we carried out parametric wavelength conversion via stimulated FWM within a triplet of coupled modes. In this process, the idler field is produced in the S supermode by tuning the pump and signal fields to the C and AS supermodes, respectively, as shown in the optimized optical spectrum in \cref{fig:dispersion-control}(f). The signal and pump fields are generated using a high-extinction EOM (40 dB) bias to suppress the optical carrier. The input (off-chip) peak power was set to 22~dBm ($3.5$~dB/facet insertion loss). For more experimental details, see Supplementary Information \cref{sec:SM_FWM}. By varying the current applied to the microheater atop ring (1), the sGVD is pretuned close to zero local dispersion regime, ensuring the phase-matching condition for stimulated FWM. 
The color code inserted into the experimental data points in \cref{fig:dispersion-control}(h) indicates the maximum power of the idler field for each sGVD condition, indicated by the value of $J_2/2\pi$. Parametric wavelength conversion is optimized close to the zero dispersion condition $(J_2 = 0)$, at which the idler power was found to be $20.4$~dB below the signal power level.
Moreover, variations of $J_2/2\pi$ of around 300~MHz can cause an efficiency loss of nearly 3~dB, and beyond that range, phase mismatch rapidly diminishes the process efficiency, and the idler becomes obscured by FWM occurring in the fibers.
Therefore, the sGVD control offered by our photonic molecule design enables optimizing or suppressing the phase-matching condition for the occurrence of FWM processes within a triplet of supermodes.

\section{DOPO generation}

\begin{figure*}[t]
    \centering
    \includegraphics[width=\linewidth]{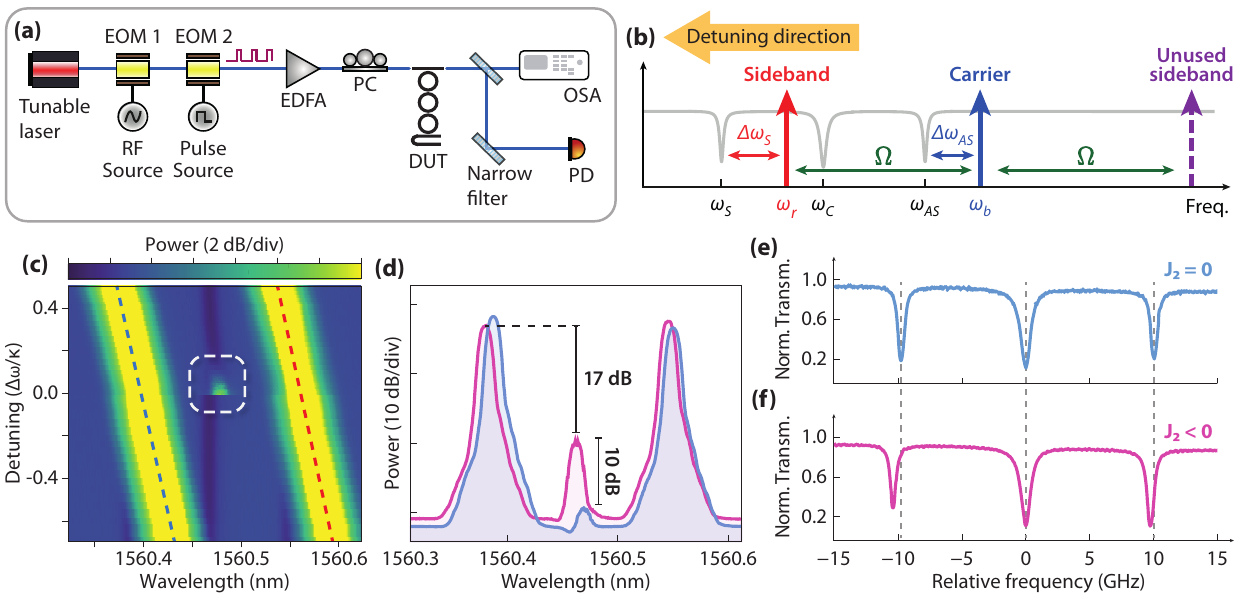}
    \caption{\textbf{DOPO measurements:} (a) Main components of the experimental setup for the DOPO measurements. Tunable laser, EOM1: electro-optic modulator driven by a sinusoidal RF-signal, EOM2: electro-optic modulator driven by a pulsed RF-signal, EDFA: erbium-doped fiber amplifier, PC: polarization controllers, DUT: device under test, OSA: optical spectrum analyzer, tunable narrow filter, and PD: photodetector. (b) Schematics of the detuning $\Delta\omega_\mathrm{AS}$ ($\Delta\omega_\mathrm{S}$) between the supermode frequency $\omega_\mathrm{AS}$ ($\omega_\mathrm{S}$) and the pump frequency $\omega_\mathrm{b}$ ($\omega_\mathrm{r}$). (c) Optical spectra collected at the through port as a function of the normalized hot detuning ($\Delta\omega/\kappa_\mathrm{C}$) between the pump fields and the photonic molecule resonances. The blue and red dashed lines indicate the pumped S and AS supermodes. The colorbar was saturated to enable the visualization of the C supermode oscillation, highlighted in white. (d) Optical spectrum at zero hot detuning of an operating DOPO for zero (blue) and normal (pink) supermode local dispersion regimes. (e,f) Transmission spectrum of the photonic molecule around 1560 nm for the zero ($J_2=0$; blue) and normal ($J_2<0$; pink) supermode local dispersion regimes.}
    \label{fig:DOPO}
\end{figure*}

DOPO is generated in the photonic molecule by pumping the outer supermodes of a triplet with a phase-coherent dual-pump scheme obtained by electro-optic modulation of a single laser. The experimental setup used to study DOPO is shown in \cref{fig:DOPO}(a).  The high-extinction EOM 1(40~dB), is biased to ensure an equal distribution of optical power between the carrier and the sidebands, and the frequency of the RF drive source is adjusted to match the frequency splitting between the AS and S supermodes ($\Omega/2\pi\approx 20$~GHz). 
The optical carrier ($\omega_\mathrm{b}$) and one of the sidebands ($\omega_\mathrm{r}$), $\omega_\mathrm{b}>\omega_\mathrm{r}$, are used to pump the supermodes $\omega_\mathrm{AS}$ and $\omega_\mathrm{S}$, as illustrated in \cref{fig:DOPO}(b).  The remaining sideband is not resonant with any resonator mode and can be neglected in our experiment.
To achieve high pump powers, necessary to surpass the DOPO threshold, the pump fields are modulated by another electro-optic intensity modulator (EOM 2), producing 6 ns pulses at a repetition rate of 1.7~MHz.
The pulse duration, approximately 17 times larger than the cavity lifetime ($\tau = 2 Q_L/\omega \approx$ 0.37~ns), is long enough to ensure a quasi-CW regime while significantly reducing thermo-optic-induced phase shifts.
The pulsed optical tones are amplified using an erbium-doped fiber amplifier (EDFA) and coupled to the chip with the aid of lensed fibers. 
From a measured input (off-chip) peak power of  8.2~W, we estimate in-chip peak power of 1.45~W per pump field.
Polarization controllers are used to ensure the excitation of the fundamental transverse electric (TE) mode. 
The output light is directed to an optical spectrum analyzer (OSA), and the DOPO signal is detected by a photodetector after filtering out the pump wavelengths (details of the filtering process can be seen at \cref{sec:SM_filter}).

DOPO occurs in the central supermode $\omega_\mathrm{C}$ as the dual-tone pumps are simultaneously red-detuned across the AS and S supermodes ($\omega_\mathrm{AS}$ and $\omega_\mathrm{S}$).
The optical spectra collected at the through port as a function of hot pump detuning are shown in \cref{fig:DOPO}(c). As the intracavity power increases, the supermode resonances are red-shifted, mainly due to self- and cross-phase modulation (SPM and XPM, respectively) contributions.
The maximum DOPO power is observed when the hot detunings of the pump fields approach zero, i.e., $\Delta\omega = \Delta\omega_\mathrm{AS}=\Delta\omega_\mathrm{S}=0$, with $\Delta\omega_\mathrm{AS}\equiv\omega_\mathrm{b}-\omega_\mathrm{AS}$ and $\Delta\omega_\mathrm{S}\equiv\omega_\mathrm{r}-\omega_\mathrm{S}$. 
Once the pump fields are detuned beyond the bistable resonance edge, the intra-cavity power abruptly drops, and the DOPO ceases to oscillate.
The unloaded supermodes' spectral positions can be observed as the darkest regions in \cref{fig:DOPO}(c) due to the broadband spontaneous emission produced by the EDFA.

In \cref{fig:DOPO}(d), the DOPO spectral signature is displayed at zero hot detuning $(\Delta\omega/\kappa_\mathrm{C}=0)$ for distinct sGVD regimes, fine-tuned with the aid of the microheater on top of one of the outer rings. 
In particular, in the normal dispersion regime ($J_2<0$, \cref{fig:DOPO}(f)), DOPO exhibits a significant improvement compared to that in the zero dispersion regime ($J_2=0$, \cref{fig:DOPO}(e)).
In the normal sGVD regime, DOPO achieves power levels 17 dB below the residual pump power, compatible with which it has been achieved in single microrings~\cite{okawachi2020demonstration, okawachi2015dual, zhao2020near}.
The interplay between the SPM and XPM effects in the supermodes plays a determinant role in the optimization of DOPO.
As the intracavity power reaches the nonlinear regime, the pumped S and AS supermodes experience SPM and XPM, while the C supermode experiences XPM contributions of the two pumps and negligible SPM. Since the effect of XPM is stronger than that of SPM, the C supermode undergoes a larger redshift compared to the S and AS supermodes, leading to an asymmetry in the frequency spacing of the triplet of coupled modes. This is compensated for by presetting the sGVD to the normal regime, as shown in \cref{fig:DOPO}(f).
In this case, the local normal dispersion imposed by the microheaters balances the nonlinear phase-shifts, thus providing an optimum phase-matching condition for DOPO.

\begin{figure}[t]
    \centering
    \includegraphics[width=.8\linewidth]{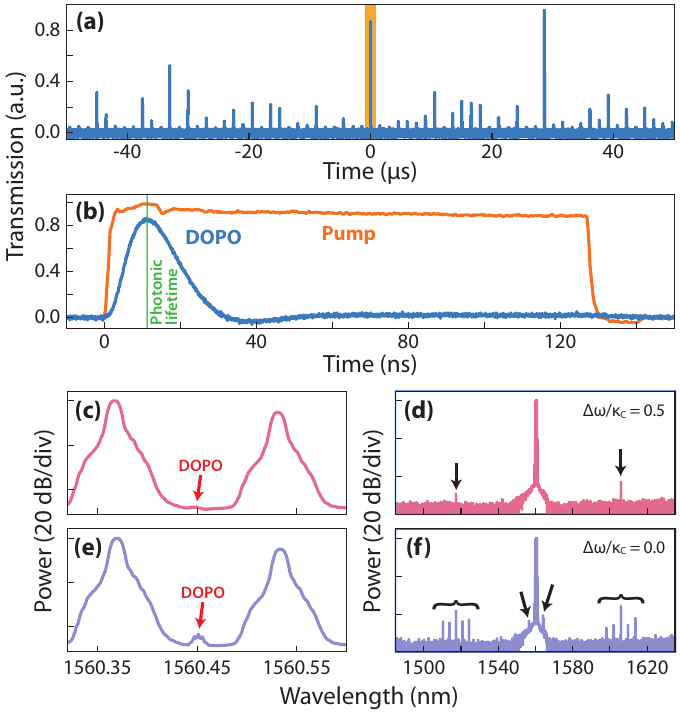}
    \caption{\textbf{Time instability and OFC:} Top (a,b) and bottom (c-f) figures present the characterization of DOPO in time and frequency, respectively. 
    (a) Transmission of the filtered DOPO signal displaying amplitude fluctuation across pulses.
    (b) Zoom-in on the highlighted DOPO pulse on (a) in blue, overlaid with the input pump pulse (127~ns, orange) and the photonic lifetime of the C supermode (0.37~ns, green).
    (c,e) Narrowband DOPO spectra centered at the triplet of interest ($\lambda_0\sim$1560~nm) at normalized hot detunings ($\Delta\omega/\kappa_\mathrm{C}$) of 0.5 and 0.0, respectively. (d,f) Broadband DOPO spectra at the same conditions of (c,e). Comb lines are indicated in black.}
    \label{fig:drawbacks}
\end{figure}

We measured the DOPO's temporal stability by filtering out the dual-tone pumps from the pulsed output light. To remove the remaining from the DOPO signal, we used a free-space Mach–Zehnder interferometer with the same FSR as the pump tones frequency spacing, cascaded by a tunable narrowband filter to get a 23 dB rejection pump-to-DOPO ratio.
For these measurements, we set the pulse duration of the dual-tone pumps to 127 ns, which also provides a quasi-CW regime within the microrings.
As the dual-tone pump is tuned towards the optical AS and S supermode resonances, oscilloscope traces reveal significant variability in the amplitude reached for each pump pulse,  as shown in \cref{fig:drawbacks}(a). Inspecting a single pump pulse, as presented in \cref{fig:drawbacks}(b), it is noticeable that the DOPO pulse is shorter than the input pump pulse, with a pulse duration of approximately 20 ns. This instability condition was verified over the whole range of pump detuning and for a range of power levels above the threshold. Our attempt to further explore the power-detuning parameter space was hindered by the excitation of spurious comb lines located far away from the pump frequencies, as illustrated in \cref{fig:drawbacks}(d,f). When the pumped power exceeds the threshold and DOPO initiates, two distant fundamental supermodes around 1517 and 1605~nm ($\mu = \pm 12$) begin to oscillate at similar power levels to DOPO itself (\cref{fig:drawbacks}(c,d)). As the dual-tone pumps further approach the AS and S resonances, more comb lines are progressively populated. Such cascading FWM occurs, despite the normal GVD regime imposed on the supermodes, because of the presence of avoided crossings between the fundamental and higher-order TE modes of the bare rings. Experimental evidence for the coupling to higher-order TE modes can be found in the fluctuations seen in the dispersion curves of \cref{fig:dispersion-control}(a-c), and in the excited comb lines beyond the triplet signature of our devices at 1517 and 1605~nm showed in \cref{sec:SM_OFC}. The suppression of higher-order modes can be achieved by exploiting design strategies that prevent their excitation by using nonlinear adiabatic couplers~\cite{kordts2016higer} or that eliminate them by using selective loss mechanisms ~\cite{ji2021exploiting} or coupling engineering \cite{PhysRevLett.127.033901}. In future work, our triple-state photonic molecule design can be further improved by incorporating these strategies. 

The instability of Kerr-based DOPO in single microresonators has been thoroughly investigated based on Lugiato-Lefever bifurcation dynamics\cite{roy2018analytical,zhao2020near, PhysRevA.90.013811}. It is known that in single microrings, the stability of the DOPO depends on the suppression of non-degenerate FWM tones in the modes near the pump wave\cite{roy2018analytical, PhysRevA.90.013811}, which are reduced by spectral engineering of the photonic molecules. Although a detailed experimental analysis of stability is needed to identify the sources of the dynamics seen in \cref{fig:drawbacks}(a,b), the temporal duration of the DOPO signal indicates that thermal effects may also be involved in our observations~\cite{gao2022probing, PhysRevApplied.20.054036}. We also do not exclude the possibility of intrinsic instability due to bifurcations inherent in the Lugiato-Lefever equations (LLEs) in coupled resonators~\cite{ghosh2024controlled}. However, our experimental findings create an opportunity to investigate the intricate behavior of dual-pumped coupled microresonators beyond the dimer regime~\cite{zhang2021squeezed}.

\section{Conclusions}

We have realized DOPO in a coupled triple-mode photonic device. In our design, the DOPO process happens entirely within the three coupled modes, without depending on modes of different azimuthal mode numbers. This helps to isolate DOPO from other nonlinear optical effects, such as non-degenerate FWM, which takes place even in the normal GVD regime. Using integrated microheaters, we could dynamically adjust the phases of the coupled modes, and thus optimize phase-matching conditions for stimulated FWM and DOPO. We have demonstrated that this active local dispersion control can be achieved independently of the overall GVD regime of the microcavities, creating more opportunities for spectral engineering. Furthermore, our design produces mode splittings with frequency difference in the RF range, enabling dual-pump schemes by electro-optic modulation of a single laser. This makes the experimental setup much easier by eliminating the need for extra lasers and amplifiers and providing native phase-locked pumps. The distant comb lines that compete with the DOPO are due to mode interactions and can be reduced with future design improvements that incorporate, for example, nonlinear adiabatic couplers \cite{PhysRevLett.127.033901}. Our triple-mode photonic device has great potential for DOPO applications, such as genuine random number generation, coherent optical computing, and generation and characterization of squeezed states of light.

\begin{backmatter}
\bmsection{Funding}
This work was supported by São Paulo Research Foundation (FAPESP) through grants 
18/15577-5, 
18/15580-6, 
18/21311-8, 
18/25339-4, 
20/04686-8, 
21/10334-0, 
22/06267-8, 
23/09412-1, 
and Coordenação de Aperfeiçoamento de Pessoal de Nível Superior - Brasil (CAPES) (Finance Code 001)

\bmsection{Acknowledgments}
The authors would like to acknowledge Prof. Felippe Barbosa for insightful discussions on the DOPO behavior.

\bmsection{Disclosures}
The authors declare no conflicts of interest.

\bmsection{Data Availability Statement}
Data underlying the results presented in this paper are not publicly available at this time but may be obtained from the authors upon reasonable request.

\bmsection{Supplemental document}
The Supplementary Material provides additional data and characterization of the DOPO presented in the main text.

\end{backmatter}

\bibliography{DOPO_photmolecule.bib}
\newpage
\renewcommand{\theequation}{S\arabic{equation}}
\renewcommand{\thesection}{S\arabic{section}}
\renewcommand{\thesubsection}{\Alph{subsection}}
\renewcommand{\thesubsubsection}{\roman{subsubsection}}
\renewcommand{\thefigure}{S\arabic{figure}}
\renewcommand{\thetable}{S\arabic{table}}
\setcounter{figure}{0}
\setcounter{table}{0}
\setcounter{equation}{0}
\setcounter{section}{0}
\section*{Supplementary Information}
\section{DOPO in photonic molecules} \label{sec:SM_DOPOinmolecules}

The slow time evolution of the supermodes' amplitudes involved in DOPO can be described by the coupled Lugiato-Lefever equations (LLE) ~\cite{herr2012universal, chembo2010modal, zeng2014design}, written in the supermode basis:
\begin{subequations}
	\begin{align}
    \frac{\partial a_{1}}{\partial\tau} &= - \left(\frac{\kappa_{1}}{\kappa_2} + i\frac{\Delta_{1}}{\kappa_2}\right) a_{1} + i\left(f^{111}_1|a_{1}|^2 + 2f^{221}_1|a_{2}|^2 + 2f^{331}_1|a_{3}|^2 \right) a_{1} + i f^{232}_1a_2^2a_{3}^\ast + s_\text{b},   \\
    \frac{\partial a_{2}}{\partial\tau} &= - \left(1 + i\frac{\Delta_{2}}{\kappa_2}\right) a_{2} + i\left(2f^{112}_2|a_{1}|^2 + f^{222}_2|a_{2}|^2 + 2f^{332}_2|a_{3}|^2 \right) a_{2} + i 2f^{123}_2a_{1}a_2^\ast a_3,   \\
    \frac{\partial a_{3}}{\partial\tau} &= - \left(\frac{\kappa_{3}}{\kappa_2} + i\frac{\Delta_{3}}{\kappa_2}\right) a_{3} + i\left(2f^{113}_3|a_{1}|^2 + 2f^{223}_3|a_{2}|^2 + f^{333}_3|a_{3}|^2 \right) a_{3} + i f^{212}_3a_2^2a_1^\ast + s_\text{r}. 
	\end{align}
    \label{eq:LLE}
\end{subequations}
The equations are normalized by $\tau = \kappa_2 t/2$, where $\kappa_j$ is the total loss of the $j^{\text{th}}$supermode. Here, the index $j=1,2,$ and $3$ is used to denote the AS, C, or S supermode, respectively. The normalized amplitude of the $j^{\text{th}}$ supermode is given by $a_j=\sqrt{2g_0/\kappa_2}A_je^{-i\Delta_jt}$, where $g_0$ is the Kerr coefficient, $A_j$ is the unnormalized amplitude of the $j^{\text{th}}$ supermode, $\omega_j$ is its frequency, and $\Delta_j$ is its detuning. 
Furthermore, $s_b$ and $s_r$ represent normalized blue and red pumps, respectively.
The third, fourth and fifth terms in the RHS of \cref{eq:LLE} denote the effects of SPM and XPM, while the sixth term accounts for the FWM interactions. Nonlinear interactions are weighted by the spatial overlap between the involved modes, calculated via the overlap integrals~\cite{chembo2010modal}:

\begin{equation}
    f^{\alpha\beta\gamma}_\mu = \frac{\omega_\mu}{\omega_0} \frac{\int \epsilon^2 \, (\vec{\mathcal{E}}^\ast_\mu \cdot \vec{\mathcal{E}}_\alpha)\,(\vec{\mathcal{E}}^\ast_\beta \cdot \vec{\mathcal{E}}_\gamma) \, dV}{\int \epsilon^2 \,||\vec{\mathcal{E}}_2||^4\,dV}.
    \label{eq:overlap}
\end{equation}
Here $\vec{\mathcal{E}}_j$ denotes the spatial mode profile of the $j^{\text{th}}$ supermode.
Notice that, due to the distinct mode distributions of the supermodes shown in \cref{fig:photonic-molecule}(e), the overlap integrals $f_\mu^{\alpha \beta \gamma}$ can not be approximated by 1. To estimate the nonlinear overlaps, we perform a change of basis to write the supermodes $\Vec{\mathcal{E}}_j(\Vec{r})$ as a linear combination of the bare modes $\Vec{e}_j(\Vec{r})$. As an example, the integrand in the numerator of the FWM coefficient $f^{123}_2$ is calculated as follows:

\begin{equation}
    \begin{split}
    \left(\Vec{\mathcal{E}}_2^\ast \cdot \Vec{\mathcal{E}}_1\right) \left(\Vec{\mathcal{E}}_2^\ast \cdot \Vec{\mathcal{E}}_3\right) 
    &= \frac{1}{8} \left(\Vec{e}_3 - \Vec{e}_1\right)^\ast \cdot \left(\Vec{e}_1 - \sqrt{2}\Vec{e}_2 +\Vec{e}_3\right)
    \left(\Vec{e}_3 - \Vec{e}_1\right)^\ast \cdot \left(\Vec{e}_1 + \sqrt{2}\Vec{e}_2 +\Vec{e}_3\right) \\
    &\approx \frac{|\Vec{e}_1|^4 + |\Vec{e}_3|^4 - 
    2 \left(\Vec{e}_1^\ast \cdot \Vec{e}_2\right)^2
    - 2\left(\Vec{e}_3^\ast \cdot \Vec{e}_2\right)^2}{8},
    \end{split}
\end{equation}
where we have discarded the field products involving rings 1 and 3 because there is no spatial overlap
between the outer rings. We chose to normalize all $f_\mu^{\alpha \beta \gamma}$ by the self-overlap of the central mode ($f_2^{222}$):

\begin{equation}
    \begin{split}
    |\Vec{\mathcal{E}}_\text{2}|^4 &= \frac{1}{4}\left[ \left(\Vec{e}_3 - \Vec{e}_1\right)^\ast \cdot \left(\Vec{e}_3 - \Vec{e}_1\right) \right]^2 \\
    &\approx \frac{1}{4} \left(|\Vec{e}_1|^2 + |\Vec{e}_3|^2 \right)^2 \\
    &\approx \frac{|\Vec{e}_1|^4 + |\Vec{e}_3|^4}{4}.
    \end{split}
    \label{eq:self-overlap}
\end{equation}

Since 
\begin{equation}
    \int \epsilon^2 \left(|\Vec{e}_1|^4 + |\Vec{e}_3|^4 \right) dV \gg 2 \int \epsilon^2 \left[2 \left(\Vec{e}_1^\ast \cdot \Vec{e}_2\right)^2 + 2\left(\Vec{e}_3^\ast \cdot \Vec{e}_2\right)^2 \right] dV,
\end{equation}
we find $f^{123}_2 \approx 1/2$. Following the same reasoning, we obtain the remaining nonlinear coupling coefficients:

\begin{equation}
    \begin{pmatrix}
    f^{111}_1 & f^{221}_1 & f^{331}_1 & f^{232}_1 \\
    f^{112}_2 & f^{222}_2 & f^{332}_2 & f^{123}_2 \\
    f^{113}_3 & f^{223}_3 & f^{333}_3 & f^{212}_3 
    \end{pmatrix}
    =
    \begin{pmatrix}
    3/4 & 1/2 & 3/4 & 1/2 \\
    1/2 & 1 & 1/2 & 1/2 \\
    3/4 & 1/2 & 3/4 & 1/2 
    \end{pmatrix}.
\end{equation}

Notice that some of the SPM and XPM terms become more prominent than the FWM terms, and XPM can no longer be described as twice as strong as SPM. At the same time, according to \cref{eq:self-overlap}, the effective mode volume of the reference mode of the molecule is twice that of the single ring, which means that the threshold power for DOPO will be higher in the molecule.

\section{Inter-ring coupling coefficient of a photonic molecule}\label{appendixA_inter-ring-coupling}  \label{sec:SM_Inter-ring coupling}
\begin{figure}[t!]
    \centering
    \includegraphics[width=\linewidth]{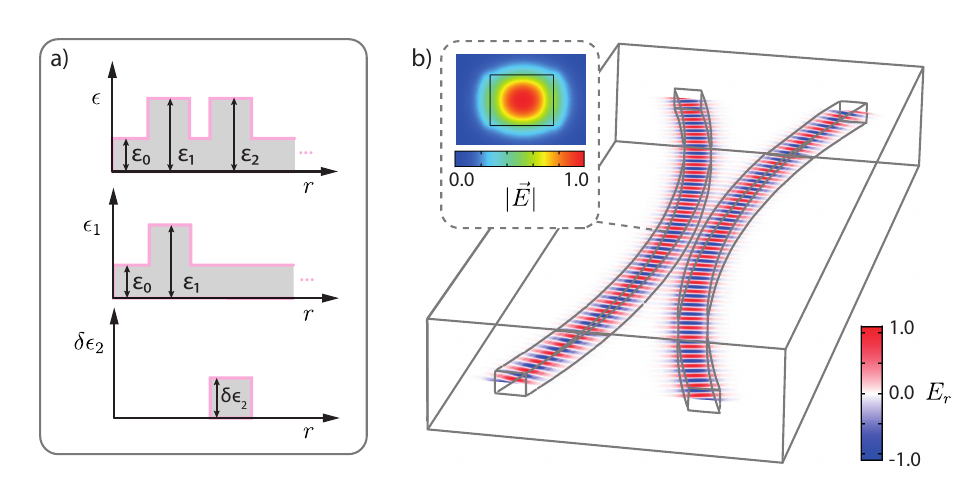}
    \caption{\textbf{Inter-ring coupling calculation:} (a) Dielectric constant distributions used in the calculation of the coupling matrix $\mathbb{H}$, represented in the transverse plane of the smallest gap between two microrings. $\epsilon$ accounts for the overall dielectric constant, $\epsilon_1$ accounts for the overall dielectric constant in the absence of the other microring, and $\delta\epsilon_2$ is the difference between $\epsilon$ and $\epsilon_1$. $\varepsilon_0$, $\varepsilon_1$ and $\varepsilon_2$ are the dielectric constants of the surrounding medium, microring 1 and microring 2, respectively. (b) 3D component enclosing the coupling domain between two microrings. The fundamental TE mode of each microring, found through a mode analysis study in Comsol, is shown in the inset. They were projected into the 3D component using Comsol extrusion operators to calculate the field overlap integrals of the matrix $\mathbb{H}$. The microrings used in this simulation are SiN microrings buried in a SiO$_{2}$ substrate with 50 $\mu$m of radius and 1 $\mu$m x 0.8 $\mu$m of width x height.}
    \label{fig:CMTcalculations}
\end{figure}

The calculation presented in this section follows a treatment of coupled mode theory (CMT) described in ref \cite{haus1991coupled} for weakly coupled and lossless microcavities. In this treatment, the electric field in the microcavity is described as a linear superposition of its mode fields: 

\begin{equation}
    \vec{E} = \sum_{i} a_{i} \, \vec{e}_{i},
	\label{eq:CMT0}
\end{equation}

\noindent in which $a_{i}$ and $\vec{e}_{i}$ are the amplitude and field profile of the $i$-th mode. Using this trial solution in Maxwell's equations, the coupled mode equations for two-coupled single-mode microcavities become: 

\begin{equation}
	\frac{d\vec{a}}{dt} = i\, \mathbb{H}\, \vec{a},
	\label{eq:CMT1}
\end{equation}

\noindent where $\vec{a} = (a_{1}\: a_{2})^{T}$ is a vector containing the mode amplitudes of each microcavity. The matrix $\mathbb{H}$ accounts for the eigenfrequencies of the individual modes (diagonal elements) and the coupling between them (off-diagonal elements). Its elements are calculated as follows: 

\begin{equation}
    	H_{ij}= - \frac{\omega_{0}}{2} \int \vec{e_i}^* \cdot \delta\epsilon_{i}\: \vec{e}_{j}\, dV,
	\label{eq:WHmatrix}
\end{equation}

\noindent where $\vec{e}_{i,j}$ and $\omega_{0}$ are, respectively, the mode field profile and angular frequency of each microcavity and $\delta\epsilon_{i} = \epsilon-\epsilon_{j}$ is the difference between the spatially-varying dielectric constant and that experienced by the $j$-th microcavity in the absence of the $i$-th microcavity. \cref{fig:CMTcalculations}(a) illustrates how the quantity $\delta\epsilon_{i}$ is calculated. The mode field profiles ($\vec{e}_{i,j}$) are normalized so that the square of their amplitudes ($|a_{i,j}|^2$) is equal to the energy carried by the modes: 

\begin{equation}
    U_{i,j} =|{\vec{a}}_{i,j}|^2\, \frac{1}{2}\,\int \epsilon |{\vec{e}}_{i,j}|^2 dV \equiv |{\vec{a}}_{i,j}|^2,
	\label{eq:CMT20}
\end{equation}

\noindent which implies:

\begin{equation}
    \vec{e}_{i,j} =\frac{\tilde{\vec{e}}_{i,j}}{\sqrt{\int\frac{1}{2} \epsilon |\tilde{\vec{e}}_{i,j}|^2 dV}}.
	\label{eq:CMT2}
\end{equation}

\noindent with $\tilde{\vec{e}}_{i,j}$ being the E-field in V/m units. 
\par The \cref{eq:CMT1,eq:CMT20,eq:CMT2,eq:WHmatrix} for a two-coupled microring system were implemented in Comsol Multiphysics. The mode field profiles ($\vec{e}_{i,j}$) of each microring were obtained through a mode analysis study performed in a 2D axisymmetric component, where the microring cross section is defined. The field profile of the TE fundamental mode of a silicon nitride (SiN) microring is shown in the inset of \cref{fig:CMTcalculations}(b).
Once the desired mode solutions were found, they were projected into a 3D component using Comsol extrusion operators. As shown in \cref{fig:CMTcalculations}(b), the 3D component encloses only the coupling domain, where the field overlap between the modes of each microring is non-zero.         


\section{Device characterization}  \label{sec:SM_characterization}

\begin{figure}[t]
    \centering
    \includegraphics[width=\linewidth]{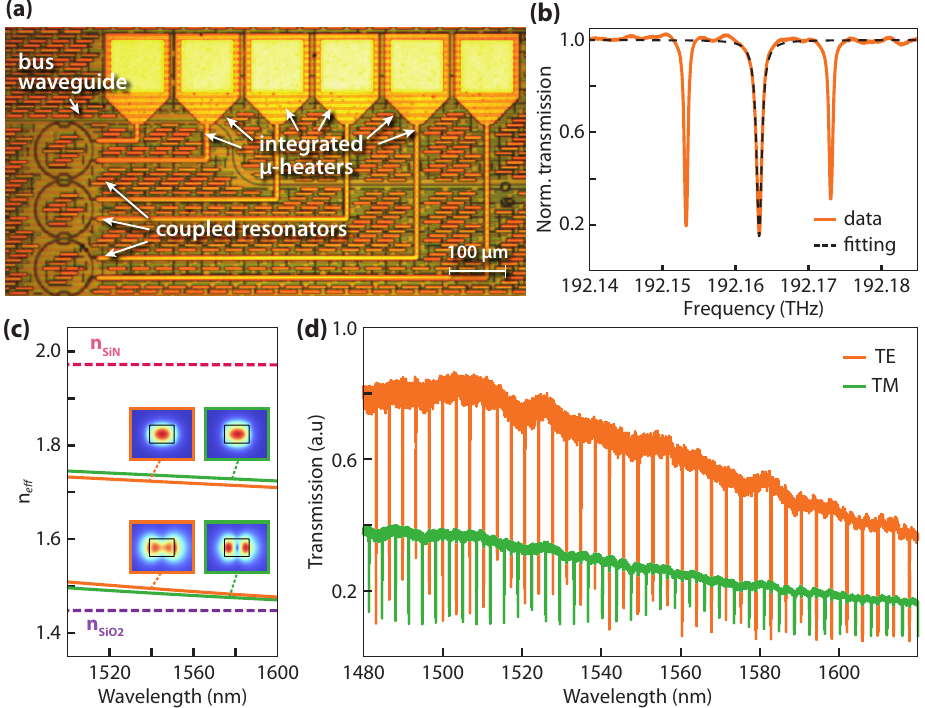}
    \caption{\textbf{Photonic molecule characterization:} (a) Micrograph of the device, showing the photonic molecule, the bus waveguide, and the integrated microheaters. (b) Transmission spectrum around the triplet of interest, used to calculate the quality factor. Individual supermodes are fitted with Lorentzian curves. The fitting of the central supermode yields a loaded quality factor of $Q_L = 2.2 \times 10^5$. (c) Effective index $n_{\text{eff}}$ as a function of the wavelength for the TE (orange) and TM (green) optical modes supported in the microring resonators employed in our photonic molecules. The fundamental (two upper curves) and first high-order modes (two lower curves) are displayed, alongside the ring waveguide (\ch{SiN}) and cladding (\ch{SiO2}) refractive indexes. The insert shows the electric field profile of each mode, simulated using the finite element method. (d) Transmission at the device's through port as the input light is polarized to excite mainly the TE (orange) and TM (green) modes.}
    \label{fig:characterization}
\end{figure}

Our photonic molecule design comprises three coupled microring resonators, illustrated in the micrograph in \cref{fig:characterization}(a). The central ring is coupled to both outer rings, each coupled solely to the central ring and a single (input or end) waveguide. Light is launched into the chip using lensed fibers, while inverse tapers optimize the coupling for TE-polarized light. The loaded Q-factor of the central supermode was found to be $Q_L = \omega_0/\Delta\omega = 2.2\times 10^5$ by fitting the narrow through-port transmission around the central supermode, shown in \cref{fig:characterization}(b), using the equation:
\begin{equation}
    T(\omega) = \frac{4(\omega-\omega_0)^2+(\kappa_i-\kappa_e)^2}{4(\omega-\omega_0)^2+(\kappa_i+\kappa_e)^2}.
\end{equation}
After determining the intrinsic and extrinsic losses, $\kappa_i$ and $\kappa_e$, respectively, the resonance linewidth $\Delta\omega$ was calculated through:
\begin{equation}
    \Delta\omega = \kappa_i+\kappa_e.
\end{equation}

Moreover, a 2D axisymmetric simulation is employed to calculate the optical modes supported by the individual microring resonators. \cref{fig:characterization}(c) shows the effective index and mode profile of the eigenmode solutions, revealing the support of the fundamental and the first higher order TE and TM optical modes.
In the photonic molecule, these modes can couple across the rings to produce higher-order or TM-polarized supermodes whose presence can potentially disturb the dispersion locally by avoided mode crossing~\cite{xue2015normal}. 
This process can occur even if the coupling between the waveguide and the photonic molecule is weaker for these parasitic supermodes. In \cref{fig:characterization}(d), we present the through-port transmission for both TE and TM polarized light, where higher-order modes are not visible. However, their presence is evident from the fluctuations in the dispersion curves of the supermodes shown in \cref{fig:dispersion-control}(a-c).

\section{General description of supermode dispersion} \label{sec:sm_supermode dispersion}

To study the dispersion of an optical resonator, the standard procedure involves expanding the mode frequencies $\omega$ around a reference mode ($m_0$) with frequency $\omega_0$, using the relative azimuthal mode $\mu = m - m_0$~\cite{fujii2020dispersion}. For photonic molecules, the dispersion can be individually calculated for each supermode by fixing the supermode index $j$:

\begin{equation}
    \omega_{j, \mu} = \omega_{j, 0} + D_{1,j}\mu + \frac{1}{2}D_{2,j}\mu^2 + \dots
\end{equation}

Here, the coefficients $D_{1,j}=\frac{\partial \omega_{j, \mu}}{\partial \mu}$ and $D_{2,j}=\frac{\partial^2 \omega_{j, \mu}}{\partial \mu^2}$ denote the average spectral spacing (FSR) and its variation in the $\mu$-space (GVD parameter). Alternatively, it might be interesting to define the supermode dispersion within a given azimuthal mode number ($\mu$ fixed)~\cite{tusnin2023nonlinear}. In a photonic molecule composed of an array of $N$ identical (degenerate) resonators, the frequency splitting induced by the linear evanescent coupling yields N-tuples that repeat for each azimuthal number $m$. Expanding the supermode frequency in relation to the supermode index results in a new dispersion relation:

\begin{equation}
    \omega_{j, \mu} = \omega_{j_0, \mu} + J_{1,j}(j-j_0) + \frac{1}{2}J_{2,j}(j-j_0)^2 + \dots
\end{equation}

In this case, $J_{1,j}=\frac{\partial \omega_{j, \mu}}{\partial j}$ represents the average frequency splitting,  $J_{2,j}=\frac{\partial^2 \omega_{j, \mu}}{\partial j^2}$ denotes its variation along the  $N$-tuple supermode, and so on. 
These define the supermode FSR (sFSR, $J_{1,j}$) and supermode GVD (sGVD, $J_{2,j}$), similar to the standard dispersion definitions in the $\mu$-space. For $N = 3$ ($j = \{\text{S, C, and AS}\}$, the design explored in this work), up to the second-order supermode dispersion is defined, which is sufficient to characterize normal ($J_{2,j}<0$) and anomalous ($J_{2,j}>0$) supermode regimes, locally within a triplet indexed by the mode number $\mu$.

\section{Four-wave mixing experimental details}  \label{sec:SM_FWM}

\begin{figure}[t!]
    \centering
    \includegraphics[width=\linewidth]{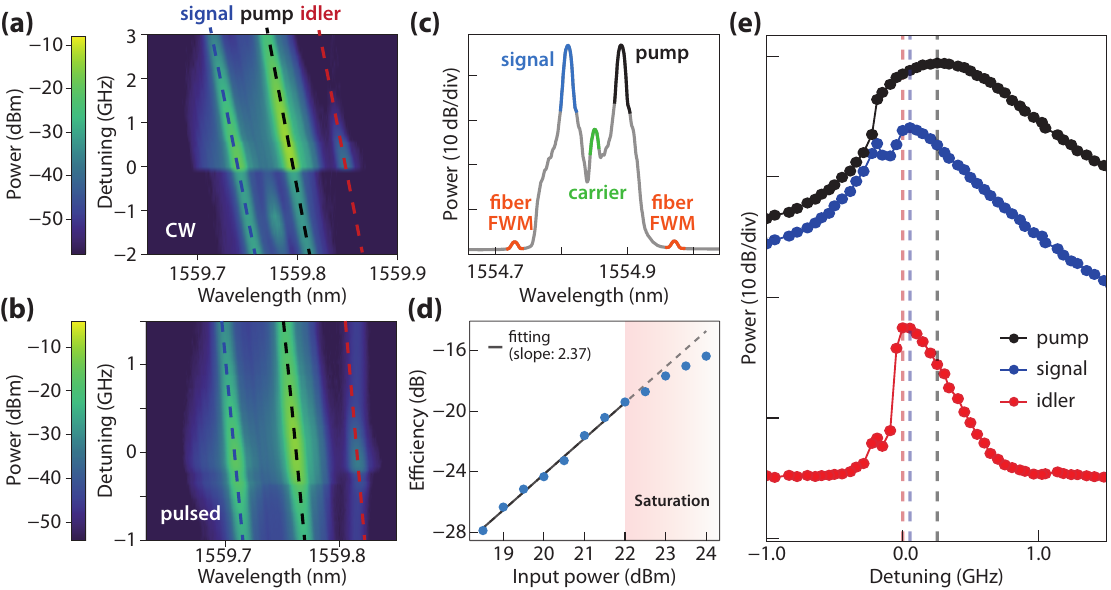}
    \caption{\textbf{FWM:}
    (a,b) FWM optical spectra collected at the drop port as a function of the hot detuning for cw (a) and pulsed (b) pumps. The dashed lines indicate the supermode frequencies, red-shifted by thermo-optic and Kerr phase shifts. 
    (c) Typical optical spectrum before the chip, showing the surpassed carrier and FWM occurring in the fibers. 
    (d) Signal-to-idler conversion efficiency as a function of the CW input power. The efficiency follows a linear trend until it saturates. 
    (e) Power of each supermode as a function of the detuning in the CW regimes. The dashed lines highlight the detuning at which the power is maximum for each curve.
    }
    \label{fig:FWM}
\end{figure}

Stimulated FWM is performed using the same setup illustrated in \cref{fig:DOPO}(a). 
We demonstrate parametric frequency conversion by adjusting the pump and signal fields across the C and AS supermodes, producing an idler field at the S supermode, as shown in \cref{fig:dispersion-control}(f).
Dual-pump tones are generated using a high-extinction EOM ($\Omega \sim 5$~GHz, half the supermode frequency splittings) biased to suppress the carrier, while an EDFA delivers 22~dBm off-chip peak power. 
Moreover, a second EOM, driven by a pulsed source, is integrated to the setup to pulse the optical pump fields, mitigating the thermo-optic effect.
\cref{fig:FWM}(c) displays a typical spectrum before the chip, showing FWM occurring in the fibers. 
In this scenario, the higher-wavelength field generated in the fiber coincides with the idler field of the photonic molecule FWM, although this does not occur in the DOPO measurements.

In \cref{fig:FWM}(a, b), the optical spectra are presented for CW (a) and pulsed (quasi-CW, b) regimes
as the dual-tone pump is simultaneously finely detuned using the laser's built-in piezo control. 
Since the pulse width of 50~ns (135 times longer than the photonic lifetime) is much faster than the thermo-optic response time~\cite{gao2022probing}, red phase-shift and bistability are drastically reduced in the pulsed regime. In \cref{fig:FWM}(d), we characterize the conversion efficiency relative to the input power. Initially, the conversion efficiency increases linearly with the pump power, but it saturates around 23 dBm due to induced phase-mismatch. 
Since the supermodes present different spatial distributions, they experience nonuniform phase-shift contributions from thermo-optic and Kerr effects. Additionally, the idler phase shift undergoes significant XPM contributions from both pumps, whereas the pumps experience SPM and XPM. Since the effect of XPM is stronger than SPM, the idler field exerts a larger red-shift.
Consequently, the supermodes achieve the bistability edge at distinct detunings, terminating the FWM process early, as shown in \cref{fig:FWM}(e).
This effect becomes more pronounced as the input power increases, affecting the efficiency trend of the process.

\section{DOPO signal filtering}  \label{sec:SM_filter}

\begin{figure}[t]
    \centering
    \includegraphics[width=\linewidth]{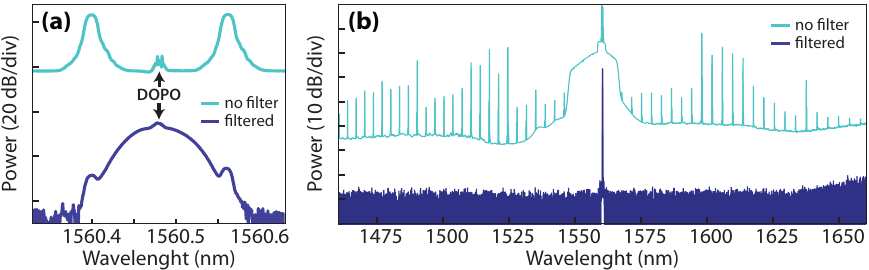}
    \caption{\textbf{DOPO filtering:} Narrow (a) and broadband (b) DOPO spectra displaying the narrow filtering achieved to isolate the DOPO signal resonance frequency. Not filtered (light blue) and filtered spectra (dark blue) are shown.}
    \label{fig:filtering}
\end{figure}

The spectral proximity of the supermodes involved in the DOPO process (separation of $\sim10~\text{GHz}$), while advantageous for the pumping scheme, introduces experimental challenges. 
This, associated with the high contrast between the two pumps and the DOPO signal, with a best-case scenario of 17 dB, complicates the filtering of the desired signal.
This challenge is particularly crucial when analyzing the temporal stability of DOPO, as shown in \cref{fig:drawbacks}(a,b). 
Even minor contributions from the pumps to the detected signal can overshadow the signal of interest when using a photodiode. 
Therefore, to enhance the relative intensity of the DOPO signal compared to the pumps, we employed a series of optical filters: first, we used a KYLIA WT-MINT delay line interferometer with variable FSR as an optical filter by setting the transmission minima to coincide with the pump positions. 
This approach was used because the interferometer is capable of withstand the high power from the chip, reducing the pump intensity to levels comparable to the DOPO signal. 
Subsequently, a Yenista XTM-50 tunable optical filter with adjustable bandwidth was used to further attenuate the pump fields and any remaining ASE from the EDFA.
As a result, the DOPO signal was enhanced to 23 dB above the filtered pumps (\cref{fig:filtering}), making it suitable for measurement with the photodiode. 

\section{Optical frequency comb}  \label{sec:SM_OFC}

Cascading FWM occurs despite the normal GVD regime imposed on the supermodes of our photonic molecule. 
Here, we show that this phenomenon is attributed to the presence of avoided crossings between the fundamental and higher-order TE modes of the bare rings. 

\begin{figure}[ht]
    \centering
    \includegraphics[width=.8\linewidth]{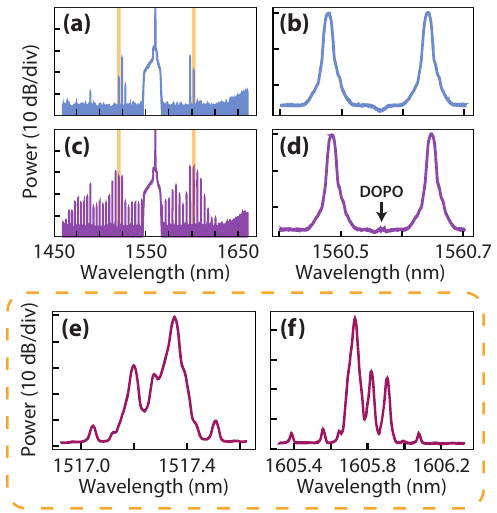}
    \caption{\textbf{OFC}
    (a) Broadband and (b) narrowband spectra collected as the pumps approach the resonances. The position of first lines spectrally distant from the pumps are highlighted in (a). The first sign of DOPO is seen in (b) as an asymmetry in the cavity resonance.
    (c) Broadband and (d) narrowband spectra collected as the detuning is reduced.
    (e, f): Zoomed view of the spectrum around the first comb lines around 1517~nm (e) and 1605~nm (f), showing excited modes beyond the triplet signature expected for the fundamental TE supermodes.}
    \label{fig:OFC}
\end{figure}

As the dual-tone pumps approach the antisymmetric (AS) and symmetric (S) resonances, comb lines are progressively populated, accompanied by the DOPO signal.
\cref{fig:OFC}(a) shows the excitation of the first comb lines around 1517 and 1605 nm ($\mu = \pm 12$) as soon as DOPO begins to oscillate. 
The first sign of DOPO is evidenced in \cref{fig:OFC}(b) by an asymmetry in the cavity resonance, seen as a result of the broadband spontaneous emission of the erbium-doped fiber amplifier.
The intensity of these initial comb lines is orders of magnitude smaller than the intensity of the pumps, indicating that the oscillation at the $\omega_c$ frequency is due to pump interaction rather than secondary cascaded nonlinear processes.
As the detuning between the pumps and AS and S resonances is further reduced, additional comb lines are populated and an increase in the DOPO amplitude is observed (\cref{fig:OFC}(c, d)). 
This dynamic process is possibly a source of the instability measured in the DOPO amplitude.

\cref{fig:OFC}(e, f) provide a zoomed view of the spectrum around the first comb lines. 
In this scenario, modes beyond the triplet signature of our devices are excited, providing experimental evidence of undesired coupling to higher-order TE modes. 
This coupling leads to avoided crossings that enable phase-matching for the initial FWM effect, causing energy to be transferred to other triplets rather than the one being pumped.

\end{document}